# Accuracy of parameterized proton range models; a comparison

H. E. S. Pettersen[a,b*], M. Chaar[b], I. Meric[c], O. H. Odland[a], J. R. Sølie[c], D. Röhrich[b]

[*]*Corresponding author: E-mail: [helge.pettersen@helse-bergen.no](helge.pettersen@helse-bergen.no).*

[a]*Department of Oncology and Medical Physics, Haukeland University Hospital, Norway*
[b]*Department of Physics and Technology, University of Bergen, Norway*
[c]*Department of Electrical Engineering, Western Norway University of Applied Sciences, Norway*

## Abstract

An accurate calculation of proton ranges in phantoms or detector geometries is crucial for decision making in proton therapy and proton imaging. To this end, several parameterizations of the range-energy relationship exist, with different levels of complexity and accuracy. In this study we compare the accuracy four different parameterizations models: Two analytical models derived from the Bethe equation, and two different interpolation schemes applied to range-energy tables. In conclusion, a spline interpolation scheme yields the highest reproduction accuracy, while the shape of the energy loss-curve is best reproduced with the differentiated Bragg-Kleeman equation.

## 1  Introduction

The proton range in well-defined geometries such as homogeneous phantoms and detector geometries can be calculated using a number of different parameterization models. Model parameters are determined through a fit to data tables from measurements of ranges in phantoms performed during commissioning and quality assessment. Several parameterizations of the range-energy relationship exist, with different levels of complexity and accuracy. For benchmarking purposes and calibration of proton range telescopes, it is important to have an accurate calculation scheme between ranges and energies. In order to calculate the pristine Bragg curve for individual proton tracks, a differentiation of the energy-range parameterization can be used.

In this study we compare the accuracy four different parameterizations models when applied in this context. High-resolution range-energy tables are created using the PSTAR database (Berger et al., 2005), and the different models are trained on the dataset. The models are then evaluated by comparing their respective errors in the reproduced ranges and the resulting Bragg curves for individual proton tracks.

## 2  Materials and Methods

In this study, different models for the relationship between the proton range and energy are evaluated based on their ability to reproduce the proton range in water at different energies as found in the Continuous Slowing Down Approximation (CSDA) from the PSTAR database.

Four models are considered: These are either semi-empirical or based on interpolation from and between data points. The semi-empirical models are derived from the Bethe equation and fitted to experimental data in order to find the parameters of the particular parameterization scheme. The interpolation-based models use different approaches to interpolate from look-up-tables of tabulated range-energy data.

The analysis is performed using ROOT 5.34/19 (Brun and Rademakers, 1997). The range-energy tables for water are calculated to 1 µm accuracy from the total stopping power as obtained from the PSTAR database. The data fitting library TMinut in ROOT has been used to find the model parameters.

## 2.1 Semi-empirical models

The Bethe equation (Bethe, 1930) describes the stopping power of charged nuclei traversing matter. Its integral is the proton range. It is not trivially integrable, however, by performing a series of approximations one obtains a simplified range-energy relationship. Several approximations have been suggested: The Bragg-Kleeman rule is the 1st term in a Taylor series, and due to its simple form, and by both inverting and differentiating the formula, one finds an expression for energy loss (Bortfeld and Schlegel, 1996). The Bragg-Kleeman rule for a proton's range $R_0$ with initial energy $E$ and energy loss curve $dE/dz$ is given as:

$$R_0 = \alpha E^p \quad (1)$$
$$E(z) = \alpha^{-1/p}(R_0 - z)^{1/p} \quad (2)$$
$$-dE/dz = p^{-1}\alpha^{-1/p}(R_0 - z)^{1/p-1}. \quad (3)$$

Here, $\alpha$ and $p$ can be obtained from the Bethe equation or from model fits to range-energy data.

Alternatively, a series of exponential terms (Ulmer, 2007) has been suggested as a more accurate model for range calculations. Two separate approximations are here offered in order to calculate $R_0$ and $E(z)$, respectively, differentiation of the latter gives rise to the $dE/dz$ curve:

$$R_0 = a_1 E_0 \left[1 + \sum_{k=1}^{N_1}(b_k - b_k \exp(-g_k \cdot E_0))\right] \quad (4)$$

$$E(z) = (R_0 - z)\sum_{k=1}^{N_2} c_k \exp(-\lambda_k(R_0 - z)) \quad (5)$$

$$-\frac{dE}{dz} = \frac{E(z)}{R_0 - z} - \sum_{k=1}^{N_2} \lambda_k c_k (R_0 - z) \exp(-\lambda_k(R_0 - z)) \quad (6)$$

The different parameters $a_1, b_k, g_k, \lambda_k$ and $c_k$ are described in (Ulmer, 2007), these are determined by fitting the model to range-energy data. A recommendation on the number of terms was also made in the same study, where $N_1 = 2$ and $N_2 = 5$ yielded a quite good accuracy. The same number of terms were therefore applied in this present work.

## 2.2 Data interpolation models

Tabulated data are used for determination of proton ranges in homogeneous phantoms of known composition. However, interpolation is needed if the required pairs of range data points are unavailable. The same holds also for more complex geometries, such as detector geometries, where the tabulated data is obtained through Monte Carlo simulations.

Linear interpolation is the simplest interpolation scheme when working with a look-up-table. Spline interpolation is performed by calculating (here) a 3rd order polynomial function around each of the data points, linking them in a piecewise fashion. It is also possible to extract energy loss curves from range-energy tables by calculating the range difference between each energy step.

## 2.3 Comparison of the parameterization models

In the present work 150 CSDA range values for protons in water, up to therapeutic energies, are split into two groups. The training group ($N_T = 25$) is used for finding the model parameters, the remaining control group ($N_C = 125$) is used to evaluate the model calculations with small range intervals: The range values are chosen from beam energies equidistantly distributed from 0 MeV to 250 MeV. Each model-calculated range is then compared to the corresponding value from the control group.

## 2.4 Comparison of the number of data values for model training

In the above analysis, the 75th percentile value of the deviation between the estimated range and the PSTAR range has been calculated for a varying number of data points used for training the different models, ranging from $N_T = 3$ to $N_T = 125$.

## 2.5 Comparison of the shape of the Bragg curve

The Bragg curve of a single proton incident on water is obtained from a differentiation of the energy-range relationship. When this is convoluted with the statistical range straggling of a proton beam and combined with the beam fluence, the result is the depth-dose curve for a proton beam, this in contrast to the pristine energy loss curve of a single proton. The different parameterizations give rise to energy loss curves with slightly different shapes.

# 3 Results

After training the models, the resulting parameters for the Bragg-Kleeman model, Eqs. 1–3, are similar to those obtained by others: see Table 1. The parameters from the fit of the sum of exponentials-model in Eqs. 4–6 are not easy to compare, due to the many terms added linearly.

The accuracy of the proton range determination using different models are shown in Figure 1 and the training stability of the models are shown in Figure 2. Energy loss curves resulting from the different models are shown in Figure 3.

|  | $\alpha$ | $p$ |
|---|---|---|
| This work | 0.00262 MeV/cm$^{-1}$ | 1.736 |
| (Bortfeld, 1997) | 0.00220 MeV/cm$^{-1}$ | 1.77 |
| (Boon, 1998) | 0.00256 MeV/cm$^{-1}$ | 1.74 |

**Table 1:** The parameters for proton range calculation using the Bragg-Kleeman rule, as found in this work and as compared with others.

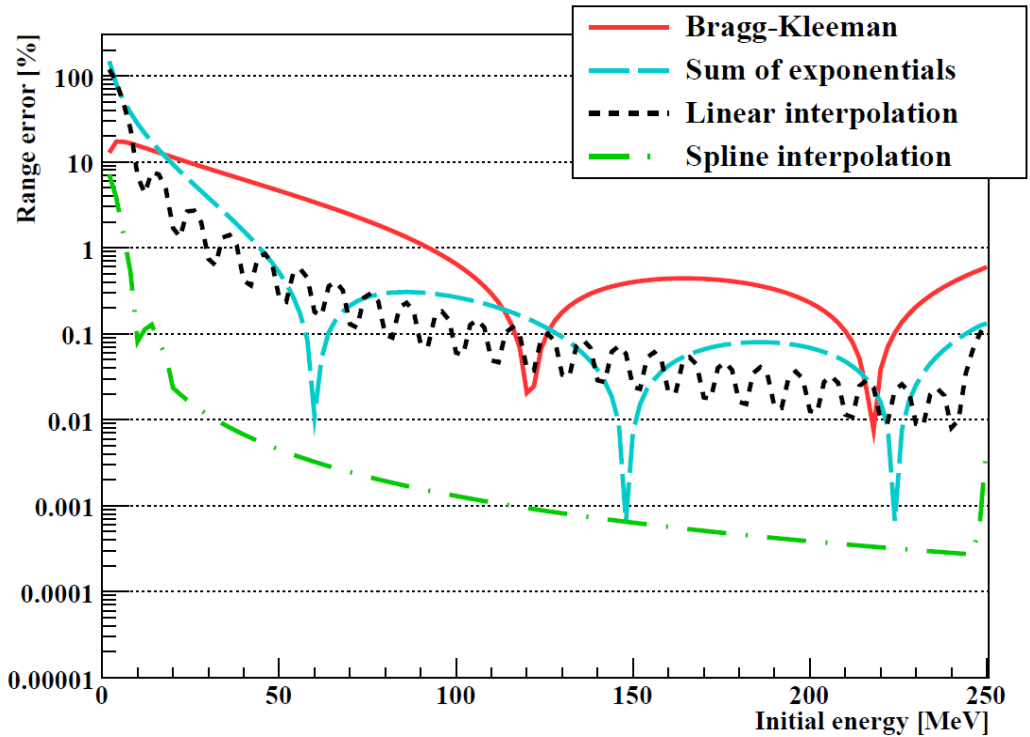

**Figure 1:** The accuracy of the proton range calculations using different parameterization models. The range error is the relative difference between the estimated range and the PSTAR data. The results are given with respect to the control group, and limited below by the PSTAR dataset accuracy of 1 μm.

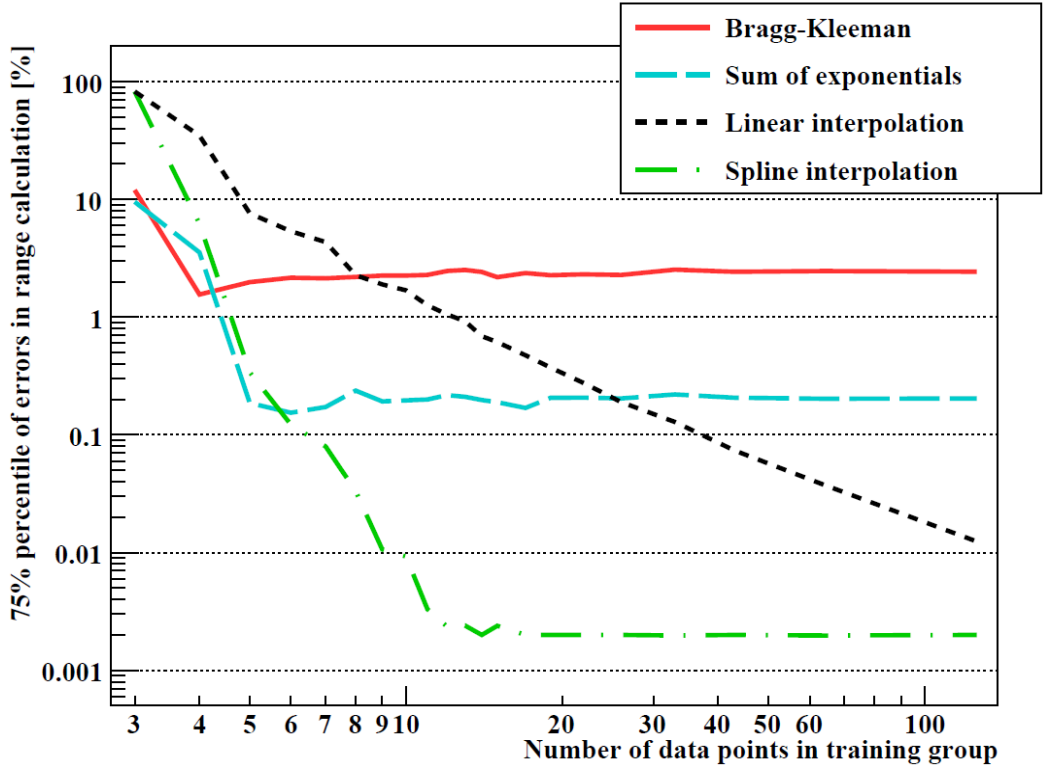

**Figure 2:** The training stability of the models as a function of the number of data points. The error shown here is calculated as the 75th percentile of all the relative errors as shown in Figure 1 for each of the parameterization models. The high accuracy of the spline interpolation is a result of its curvature.

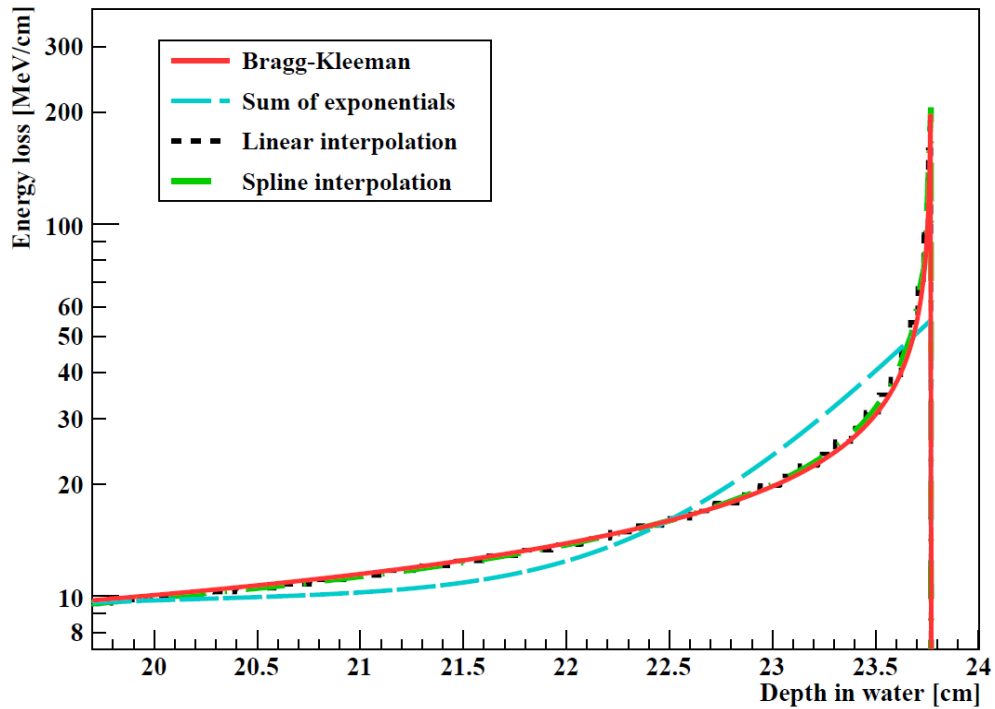

**Figure 3:** The energy loss curves for individual 190 MeV protons obtained by differentiating the parameterization models. The range is kept constant by choice of $R_0$, as to facilitate a comparison between the curve shape.

# 4 Discussion and conclusion

The spline interpolation model yields the highest accuracy. A sub-percent range calculation accuracy is obtained for all models above 90 MeV, and for the spline interpolation model above 10 MeV.

A larger number of measurements at different energies are required for an interpolation-based range calculation scheme compared to the simple Bragg-Kleeman rule with two parameters, or the exponential sum with five parameters. Using 25 data points for training the model, the accuracy is kept at an acceptable level for all models, and the 75th percentile of the errors in the range calculation is at 0.1% of the range for both interpolation schemes and the sum of exponentials.

The shapes of the Bragg curves obtained from the interpolations and the Bragg-Kleeman model are identical. The curves produced by the interpolation methods are stepwise functions since the number of data points are limited. The curve obtained by using the sum of exponentials exhibits differences close to the Bragg Peak mimicking an exponential decay. In conclusion the shape and the stopping position of the Bragg curve of individual protons is accurately represented by using the simple differentiated Bragg-Kleeman formula combined with a range calculated by spline interpolations.

An application for this work is found in the range calculations for proton range telescopes (Pettersen et al., 2017). A look-up-table of range-energy values is created using Monte Carlo (MC) simulations used in conjunction with spline interpolations. The experimental and MC calculated Bragg curves for individual protons are there compared to those originating from the differentiated Bragg-Kleeman formula. The result is high accuracy of both range calculation of arbitrary energies as well as quite realistic parametric Bragg curves for individual protons.

## Acknowledgements
This project was supported by Helse Vest RHF [911933].